\documentstyle[11pt,aaspp4]{article}

\slugcomment{Appears in {\it The Astrophysical Journal}, {\bf543}, 787, 2000.}

\lefthead{Lay and Halverson}
\righthead{Atmospheric Fluctuations and Degree-Scale Imaging}

\begin{document}

\title{The Impact of Atmospheric Fluctuations on Degree-scale Imaging 
of the Cosmic Microwave Background}

\author{Oliver P. Lay}
\affil{Radio Astronomy Laboratory, University of California, Berkeley,
CA 94720}

\and

\author{Nils W. Halverson\altaffilmark{1}} 
\affil{Dept.\ of Astronomy and Astrophysics, University of Chicago, 
5640 S.\ Ellis Avenue, Chicago, IL 60637}

\altaffiltext{1}{Dept.\ of Applied Physics, 
California Institute of Technology, Pasadena, CA 91125}

\begin{abstract}

Fluctuations in brightness due to water vapor in the Earth's
atmosphere are an important source of noise for ground-based
instruments attempting to measure the anisotropy of the Cosmic
Microwave Background. This paper presents a Kolmogorov model of
atmospheric fluctuations, and derives simple expressions to predict
the impact of fluctuations on experimental measurements for three
instrument configurations: chopped beam, swept beam and
interferometer. Data from the South Pole and from the Atacama Desert
in Chile, two of the driest places on Earth, are used to characterize
the fluctuations at each site. Using an interferometric instrument as
an example, the data suggest that the South Pole is the superior site
for observations of the Cosmic Microwave Background at degree angular
scales.

\end{abstract}


\section{Introduction}

Many groups are currently engaged in measuring the level of anisotropy
present in the cosmic microwave background (CMB) from arcminute to
degree angular scales. The brightness temperature of the fluctuations
is of order $10^{-5}$~K, requiring carefully designed
experiments. Contamination from point sources and galactic dust
emission is minimized by choosing an observing frequency of between 20
and 300~GHz (1.5~cm to 1~mm), depending on the angular scale being
investigated (Tegmark \& Efstathiou 1996).  Differential measurements
are employed to minimize systematic errors: chopped beam instruments
measure the difference in emission between two or more directions on
the sky, swept beam instruments sweep a beam rapidly backwards and
forwards, and interferometers correlate signals from two or more
antennas. Some examples of chopped beam experiments are Python I-IV
(Dragovan et al. 1994, Ruhl et al. 1995, Platt et al. 1997, Kovac et
al. 2000), OVRO RING5M (Leitch et al. 1998), and Tenerife (Davies et
al. 1996). Current or recent swept beam experiments include Python V
(Coble et al. 1999), Saskatoon (Netterfield et al. 1997), and the Mobile Anisotropy Experiment (Torbet et al. 1999). Interferometers currently being built include the
Degree Angular Scale Interferometer (DASI) (Halverson et al. 1998), the Cosmic Background
Interferometer (CBI) (A.C.S Readhead \& S. Padin, personal communication), and the Very Small Array (VSA) (Jones \& Scott 1998). In addition, the
Cosmic Anisotropy Telescope (CAT) has already been operational for
several years (Scott et al.\ 1996). These employ wide bandwidth,
low-noise receivers to maximize sensitivity, and should be deployed at
sites where the Earth's atmosphere does not significantly compromise
performance.

The atmosphere is also a source of brightness temperature variations,
originating primarily from water molecules. Of the the three states
that may be present -- vapor, liquid and ice -- it is water vapor that
is most important. Most of it is contained in the troposphere with a
scale height of $\sim 2$~km, and, because the water vapor is close to
its condensation point, it is poorly mixed with the `dry' component of
the atmosphere (mostly nitrogen and oxygen). This, in combination with
turbulence, leads to a clumpy, non-uniform distribution of water vapor
in the troposphere. Since the water molecule has a strong dipole
moment, rotational transitions couple strongly to millimeter-wave
radiation, and water vapor is the dominant source of atmospheric
emission (and therefore opacity) at most millimeter
wavelengths. Liquid water, in the form of clouds, is also a source of
non-uniform emission, but radiates much less per molecule. Ice is the
least efficient radiator, since the molecules are unable to
rotate. Fluctuations in temperature caused by turbulent mixing are
also a source of brightness temperature variations, although they are
generally much less significant than the water vapor contribution at
millimeter wavelengths.

It should also be noted that the high refractive index of water vapor
causes an excess propagation delay, and a non-uniform distribution of
water vapor distorts an incoming wavefront. This sets a `seeing' limit
on interferometric observations at millimeter wavelengths, and there
is an ongoing effort to correct for this effect. It is this problem,
which limits performance at arcsecond spatial resolution and high
frequencies, that has driven much of the recent research into the
distribution of water vapor (e.g.~Armstrong \& Sramek 1982; Treuhaft
\& Lanyi 1987, Wright 1996, Lay 1997).

The wavefront distortions are not significant for the low resolution
experiments considered here. The atmospheric emission fluctuations,
however, can only be distinguished from fluctuations in the CMB by the wind-induced motion of the atmosphere with respect
to the background. This paper investigates how well the two can be
separated. The next section describes a model of atmospheric
fluctuations and the responses of the different types of instrument.
Section~\ref{python} describes how data from the Python~V experiment
have been used to characterize the fluctuations at the South Pole, and
Section~\ref{chile} estimates the level of emission fluctuations for
the Atacama Desert in Chile using rms path fluctuation data.
Section~\ref{example} shows how the theory of
Section~\ref{models} can be combined with the fluctuation data from
each site to predict the residual noise level due to the atmosphere for a given instrument configuration.

\section{Models}
\label{models}

In this section, we develop a model that describes the atmospheric 
fluctuations and their interaction with the instruments commonly used to 
measure CMB fluctuations.

Church~(1995) presented a rigorous mathematical analysis based on the 
autocorrelation function of the fluctuations. We adopt a more pictorial approach based on the
power spectrum of the fluctuations, using an atmospheric model that
differs in a fundamental way from that assumed by Church.

\subsection{Sky brightness and instrument response}
\label{framework}

A given pointing on the sky can be represented by a point on the surface of a sphere. Consider a patch of sky with angular extent $\Delta\theta_x \ll 1$~radian, $\Delta\theta_y \ll 1$~radian, such that the curved surface is well represented by a pair of ortholinear angular coordinates $(\theta_x, \theta_y)$. In addition, we will consider only the fine structure (angular scales $\ll 1$~radian) in the sky brightness distribution. These approximations greatly simplify the analysis; simple Fourier Transforms can be used, and the gradient in brightness temperature due to airmass can be ignored, without significant impact on the degree scales of interest.

The Fourier Transform of the (fine structure) sky brightness distribution $T_{\rm sky}(\theta_x,\theta_y)$ is given by 
\begin{equation}
\tilde{\bf T}_{\rm sky}(\alpha_x, \alpha_y) = 
\frac{1}{\sqrt{\Delta\theta_x \Delta\theta_y}}
\int_{-\Delta\theta_x/2}^{+\Delta\theta_x/2} 
\int_{-\Delta\theta_y/2}^{+\Delta\theta_y/2} 
T_{\rm sky}(\theta_x, \theta_y) e^{-2\pi i(\alpha_x\theta_x+\alpha_y\theta_y)} 
d\theta_x d\theta_y. 
\label{T transform}
\end{equation}
The angular wavenumbers $(\alpha_x,\alpha_y)$ have units of cycles per radian. In general $\tilde{\bf T}_{\rm sky}(\alpha_x, \alpha_y)$ is a complex variable. Its amplitude is denoted by $\tilde{T}_{\rm sky}(\alpha_x, \alpha_y)$. The normalization in equation~(\ref{T transform}) is chosen such that the variance of the sky brightness fluctuations is given by
\begin{eqnarray}
T_{\rm sky,rms}^2 &=& 
\frac{1}{\Delta\theta_x \Delta\theta_y}
\int_{-\Delta\theta_x/2}^{+\Delta\theta_x/2} 
\int_{-\Delta\theta_y/2}^{+\Delta\theta_y/2} 
T_{\rm sky}^2(\theta_x, \theta_y) d\theta_x d\theta_y \\
&=& \int_{-\infty}^{+\infty} 
\int_{-\infty}^{+\infty} 
\tilde{T}_{\rm sky}^2(\alpha_x, \alpha_y) d\alpha_x d\alpha_y,
\end{eqnarray}
where $\tilde{T}_{\rm sky}^2$ represents the Power Spectral Density (PSD) of the (fine structure) sky brightness distribution (units: temperature$^2$~radian$^2$).

A radiometer has a beam pattern represented by gain function $G(\theta_x^\prime, \theta_y^\prime)$, where $\theta_x^\prime$ and $\theta_y^\prime$ are angular offsets from the optical axis of the instrument. For a beam pattern localized within $\theta_x^\prime \ll 1, \theta_y^\prime \ll 1$, the response to fluctuations of a given angular wavenumber is
\begin{equation}
\hat{\bf G}(\alpha_x, \alpha_y) =
\int_{-\infty}^{+\infty} 
\int_{-\infty}^{+\infty} 
G(\theta_x^\prime, \theta_y^\prime) 
e^{-2\pi i(\alpha_x\theta_x^\prime+\alpha_y\theta_y^\prime)} 
d\theta_x^\prime d\theta_y^\prime. 
\label{G transform}
\end{equation}
The gain $G(\theta_x^\prime, \theta_y^\prime)$ is normalized such that the maximum value of $\hat{G}(\alpha_x,\alpha_y)$ is unity. The hat symbol is used instead of the tilde to distinguish between the different normalizations applied in equation~(\ref{T transform}) and equation~(\ref{G transform}). The response of the instrument to the sky is given by the convolution of the beam and the sky brightness. In the angular wavenumber domain,
\begin{equation}
\tilde{\bf T}_{\rm inst} = \tilde{\bf T}_{\rm sky} \hat{\bf G}.
\end{equation}
The instrument acts as a spatial filter that is only sensitive to a particular range of angular wavenumbers on the sky. For example, a wide-angle beam smears out the small-scale structure, and the corresponding $\hat{G}$ has a narrow distribution. An ideal pencil beam responds equally to all angular wavenumbers with $\hat{G}=1$.   

Atmospheric features in the sky brightness distribution are blown across the angular coordinate frame, causing the output of the radiometer to vary with time. Features on the sky with large angular wavenumber (many cycles per radian) produce rapid variations with time compared to those with small angular wavenumber. Time averaging of the instrument output over a period $t_{\rm av}$ suppresses the rapid fluctuations (large wavenumber). This relationship can be expressed by including a temporal filter function $\hat{Q}(\alpha_x, \alpha_y)$, i.e.
\begin{equation}
T^2_{\rm out,rms} = 
\int_{-\infty}^{+\infty}
\int_{-\infty}^{+\infty} 
\tilde{T}^2_{\rm sky} \hat{G}^2 \hat{Q}^2 d\alpha_x d\alpha_y.
\label{filters}
\end{equation}
   
Equation~(\ref{filters}) can be used to calculate the rms noise due to atmospheric emission fluctuations for a radiometric system with time averaging. 
Section~2.2 describes an approximate model for $\tilde{T}_{\rm sky}^2$. 
Section~2.3 derives the spatial filter function $\hat{G}$ for
chopped beam, swept beam and interferometric experiments. 
Section~2.4 characterizes the instrument temporal response $\hat{Q}$. Section~2.5 applies the results of 2.2, 2.3 and 2.4 to derive the residual level of
atmospheric fluctuations for a measurement, using equation~(\ref{filters}). 
Section~2.6 highlights the atmospheric parameter most relevant for degree-scale imaging of the CMB.

\subsection{Model of Atmospheric Emission Fluctuations}
\label{Kolmogorov}

We adopt the Kolmogorov model of turbulence (Tatarskii
1961), summarized briefly below. Turbulent energy is injected into the atmosphere on large
scales from processes such as convection and wind shear, and then
cascades down through a series of eddies to smaller scales, until it is 
dissipated by viscous forces on size
scales of order 1~mm. If energy is conserved in the cascade then
simple dimensionality arguments can be used to show that the power
spectrum of the fluctuations in a large 3-dimensional volume is
proportional to $q^{-11/3}$, where $q$ is the spatial wavenumber
(units:~length$^{-1}$). This holds from the outer scale size $L_{\rm
o}$ on which the energy is injected to the inner scale size $L_{\rm
i}$ on which it is dissipated. Tatarskii showed that this same power
law applies to quantities that are passively entrained in the flow of
air, such as the mass fraction of water vapor.

\begin{figure}
\begin{center}
\leavevmode
\epsfxsize=0.45\columnwidth
\epsfbox{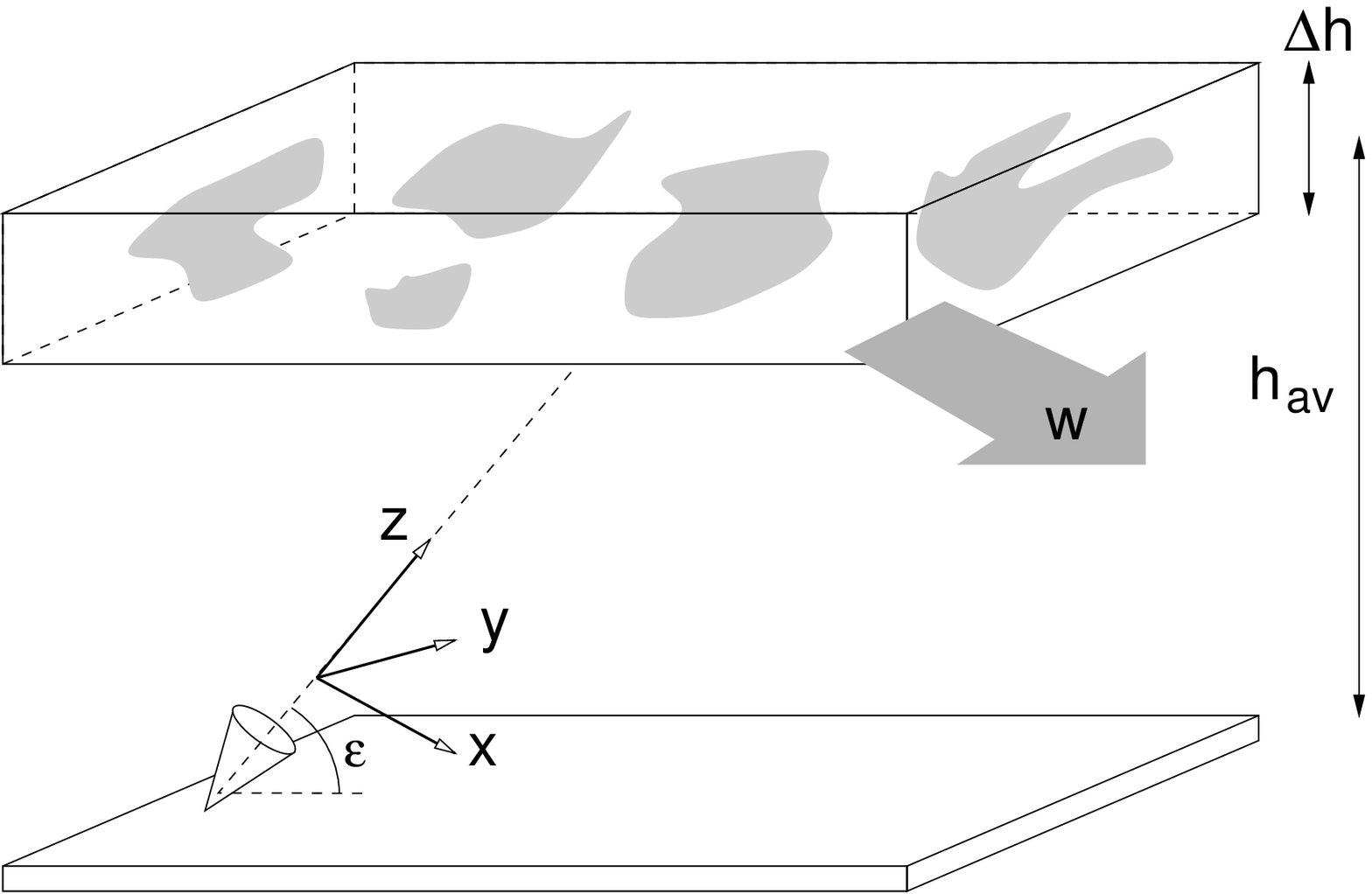}
\end{center}
\caption{Geometry adopted for the analysis. The generic instrument is 
observing at elevation $\epsilon$ through a layer of fluctuations with
thickness $\Delta h$ at average altitude $h_{\rm av}$ that is being
blown by wind vector ${\bf w}$. The $x$-, $y$- and $z$-axis form an
orthogonal set with the $z$-axis parallel to the line of sight, and
the $x$-axis parallel to the projected wind
direction.\label{geometry}}
\end{figure}

Figure~\ref{geometry} shows the geometry used for this analysis. The
water vapor fluctuations are present in a layer of thickness $\Delta
h$ at average altitude $h_{\rm av}$. The $x$-, $y$- and $z$-axes form
an orthogonal set with the $z$-axis parallel to the line of sight of
the observations at elevation $\epsilon$. 

Consider first the case of observing in the zenith direction,
$\epsilon=90^\circ$.  In the optically thin limit, the brightness
temperature contribution at frequency $\nu$ from a volume element of
atmosphere with thickness $dz$ along the line of sight and water vapor
density $\rho_{\rm H_2O}$ is given by
\begin{equation}
dT_{\rm sky}(x,y,z,\nu) = \rho_{\rm H_2O}(x,y,z) 
\kappa_{\rm H_2O}(\nu) T_{\rm phys}(x,y,z) dz,
\end{equation}
where $\kappa_{\rm H_2O}(\nu)$ is the mass opacity function for water
vapor and $T_{\rm phys}$ is the physical temperature of the volume
element. The brightness temperature of the atmosphere
looking vertically up from the ground is
\begin{equation}
T_{\rm sky}(x,y,\nu) = \kappa_{\rm H_2O}(\nu) 
                       \int_0^\infty T_{\rm phys}(z) \rho_{\rm H_2O} dz.
\end{equation}
The dependence on $\nu$ is considered to be implicit from now on. The
Fourier transform of this distribution is
\begin{equation}
\tilde{T}_{\rm sky}(q_x, q_y) = \lim_{X,Y\rightarrow\infty}\left[\frac{1}{\sqrt{XY}}
\int_{-Y/2}^{+Y/2}\int_{-X/2}^{+X/2}T_{\rm sky}(x,y)e^{-2\pi i(q_x x + q_y y)} dx dy
\right],
\end{equation}
where $q_x$ and $q_y$ are the spatial wavenumbers for the $x$ and $y$
directions. 

For Kolmogorov turbulence it can be shown (e.g.~Lay 1997) that the Power Spectral Density (PSD) for the fluctuations as seen in projection is given by
\begin{equation}
\langle\tilde{T}_{\rm sky}^2(q_x,q_y)\rangle = 
\cases{
A (q_x^2+q_y^2)^{-11/6} 
\quad\quad L_{\rm i} \ll |q_x^2+q_y^2|^{-1/2} \ll 2\Delta h \cr
A^\prime (q_x^2+q_y^2)^{-8/6} 
\quad\quad 2\Delta h \ll |q_x^2+q_y^2|^{-1/2} \ll L_{\rm o}
} 
\label{erae}
\end{equation}
The coefficient $A$, and the related value $A^\prime$, are a measure
of the turbulent intensity.  $\langle\tilde{T}_{\rm sky}^2\rangle$ has units of
K$^2$~m$^2$ and the angular braces denote an average over many realizations of the random atmosphere. 
The first case in equation~(\ref{erae}), applies to spatial wavelengths smaller
than $2\Delta h$, where the turbulence is considered to be isotropic in
three dimensions. In the second case the turbulence is isotropic in
the horizontal plane, but is constrained vertically to lie in a layer
of thickness $\Delta h$. The increase of the PSD as $(q_x,q_y)
\rightarrow (0,0)$ is less rapid than for the 3D case. Beyond the
outer scale $L_o$ the PSD should become constant.

\begin{figure}
\begin{center}
\leavevmode
\epsfxsize=0.45\columnwidth
\epsfbox{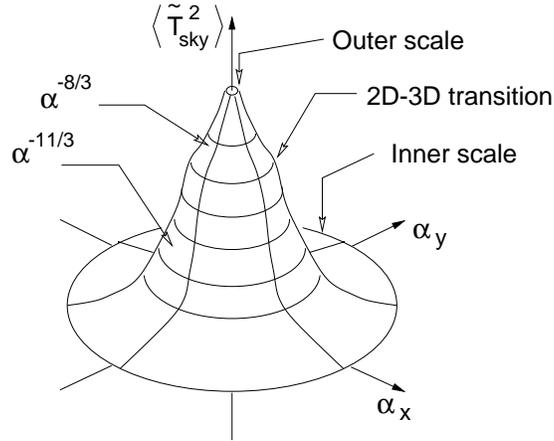}
\end{center}
\caption{Schematic representation of the distribution of atmospheric 
fluctuation power in the angular wavenumber domain. The combination
$\alpha_x = 10$, $\alpha_y = 0$ would correspond to a fluctuation with
a periodicity of 0.1~radian ($5.7^\circ$) in the $x$-direction. In
practice the distribution is distorted within
$(\alpha_{x}^2+\alpha_{y}^2)^{1/2} \sim 1$, since the angular
approximation breaks down.\label{angular_power}}
\end{figure}

The derivation of the $-11/3$ power law by Kolmogorov applies in the
isotropic three-dimensional case, well within the inner and outer
scales, where the turbulence can be considered scale-free. There is
extensive experimental evidence to support this.  In previous analyses
(e.g.~Church 1995, Andreani et al.~1990), it was assumed that there
were no correlations present in the turbulent layer on scales greater
than the thickness $\Delta h$, i.e.~$\Delta h$ corresponded to the
outer scale size (outer scale sizes of between 1~m and 100~m were used
in Church's calculations). Data from atmospheric phase monitors
operating at 12~GHz (Masson~1994; Holdaway~1995, Lay~1997) and radio
telescope arrays such as the Very Large Array (Armstrong \& Sramek
1982; Carilli \& Holdaway 1997) show no evidence for such a low value
of $L_{\rm o}$; indeed, correlation is observed over separations in
excess of 10~km.  While there may be conditions in which the outer
scale length is greatly reduced (Coulman \& Vernin 1991), the data are
generally well described by a $-11/3$ power law on small scales, and a
$-8/3$ power law on large scales, with the transition occurring for
sizes comparable to the layer thickness as described in
equation~(\ref{erae}). The $-8/3$ power law for the two-dimensional regime
can be derived from similar scaling arguments to the 3D case, but
cannot be justified well on theoretical grounds, since
turbulence is not strictly possible in a two-dimensional medium -- the
layer must have some vertical extent -- and the model should therefore
be regarded as somewhat empirical. This
distribution of fluctuation power is shown schematically in
Fig.~\ref{angular_power}.

Linear coordinates $(x,y)$ for fluctuations in a layer at average
height above the ground, $h_{\rm av}$, are converted to angular
coordinates in radians: $\theta_x \simeq x/h_{\rm av}, \theta_y \simeq
y/h_{\rm av}$ (still assuming $\epsilon\sim 90^\circ$). The approximation holds when $\theta_x \ll 1$ and
$\theta_y \ll 1$. Linear wavenumbers $(q_x,q_y)$ are converted to
angular wavenumbers: $\alpha_x = q_x h_{\rm av}$, $\alpha_y = q_y
h_{\rm av}$. Therefore
\begin{equation}
\langle\tilde{T}_{\rm sky}^2(\alpha_x,\alpha_y)\rangle = 
\cases
{
A h_{\rm av}^{5/3}(\alpha_x^2+\alpha_y^2)^{-11/6} 
\quad\quad h_{\rm av} / (2\Delta h) \ll (\alpha_x^2+\alpha_y^2)^{1/2} \ll \alpha_{\rm i} \cr
A^\prime h_{\rm av}^{2/3} (\alpha_x^2+\alpha_y^2)^{-8/6} 
\quad\quad \alpha_{\rm o} \ll (\alpha_x^2+\alpha_y^2)^{1/2} \ll h_{\rm av} / (2\Delta h). 
}
\label{sdge}
\end{equation}
The transition between the three- and two-dimensional regimes is at
approximately $h_{\rm av} / (2\Delta h)$; $\alpha_{\rm o}$ and
$\alpha_{\rm i}$ correspond to the outer and inner scales of the
turbulence, respectively. A factor of $h_{\rm av}^2$
is included for the correct normalization of the PSD, such that $\langle\tilde{T}_{\rm sky}^2 (\alpha_x,\alpha_y)\rangle d\alpha_x d\alpha_y = \langle\tilde{T}_{\rm sky}^2 (q_x,q_y)\rangle dq_x dq_y$.

Up until this point it has been assumed that the observations are in
the zenith direction.  In the 3D regime, two factors are needed to
scale to arbitrary elevation. First, the atmospheric fluctuation
power is proportional to the path length through the layer, which
scales as $1/\sin\epsilon$, i.e.~$A$ in equation~(\ref{sdge}) must be
replaced by $A/\sin\epsilon$. In addition, the distance to the layer
of fluctuations has increased from $h_{\rm av}$ to $h_{\rm
av}/\sin\epsilon$. Applying these substitutions to equation~(\ref{sdge}), for
the 3D regime only ($h_{\rm av}/(2\Delta h\sin\epsilon) \ll
(\alpha_x^2+\alpha_y^2)^{1/2} \ll \alpha_{\rm i}$)
\begin{eqnarray}
\langle\tilde{T}_{\rm sky}^2(\alpha_x,\alpha_y)\rangle &=& 
\left(\frac{A}{\sin\epsilon}\right) 
\left(\frac{h_{\rm av}}{\sin\epsilon}\right)^{5/3} 
(\alpha_x^2+\alpha_y^2)^{-11/3} \\
&=& A h_{\rm av}^{5/3} (\sin\epsilon)^{-8/3} \alpha_{xy}^{-11/3}.
\end{eqnarray}
This prescription for atmosphere fluctuations, valid for small angular
scales, is used in \S\ref{residual}. On larger scales, where the 2D
regime becomes more important, the situation has a much more
complicated dependence on layer thickness and elevation, and a
numerical calculation should be performed (e.g.~Treuhaft and Lanyi
1987; Lay 1997).

\subsection{Spatial Filtering}

This section is concerned with the derivation of the spatial filter function $\hat{G}^2(\alpha_x,\alpha_y)$ (eq.~[\ref{filters}]) for three common instrument configurations: (1)~chopped beam, (2)~swept beam, and (3)~interferometer. The function $\hat{G}(\alpha_x,\alpha_y)$ is the Fourier Transform of the instrument gain pattern, $G(\theta_x,\theta_y)$ (eq.~[\ref{G transform}]).
We assume that the emission fluctuations are present in the far-field of the instrument, i.e.~$h_{\rm av} \gg \frac{D^2}{\lambda}$, so that this analysis does not apply to large telescopes and short wavelengths. 

The result for a single aperture with a circular, Gaussian beam pattern is useful for all 3 configurations. For a beam with Full-Width-to-Half-Maximum (FWHM) power of $\theta_{\rm b}$, it can be shown that the spatial filter is another circular Gaussian:
\begin{equation}
\hat{G}^2(\alpha_x,\alpha_y) = \exp\left\{-\frac{\pi^2\theta_{\rm b}^2 (\alpha_x^2+\alpha_y^2)}{2\log_e 2}\right\},
\end{equation}
with FWHM of $0.62\theta_{\rm b}^{-1}$.

\subsubsection{Chopped beam}

\begin{figure*}
\begin{center}
\leavevmode
\epsfxsize=0.75\columnwidth
\epsfbox{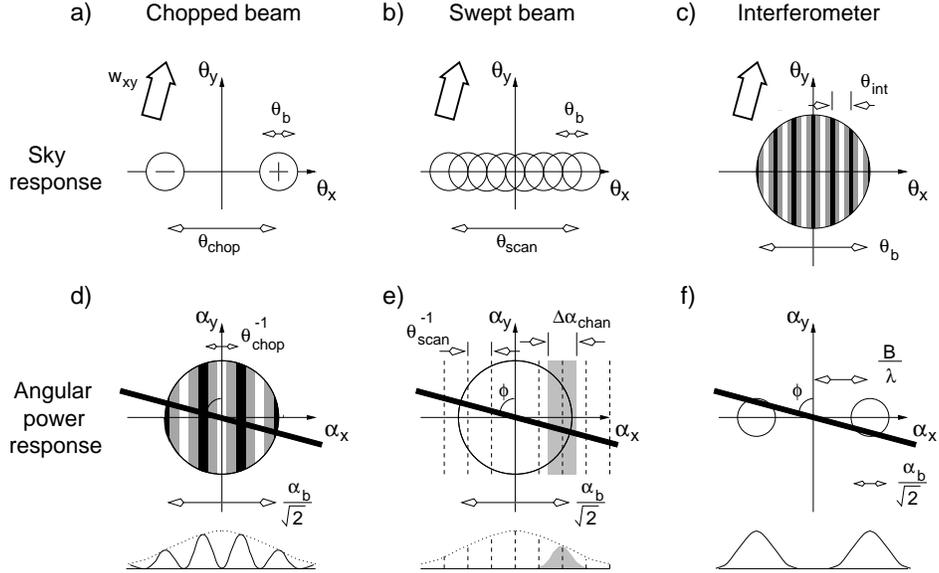}
\end{center}
\caption{Schematic responses of a chopping single aperture (a), a swept 
beam single aperture (b) and an interferometer (c) to far-field
emission on the sky. The square of the Fourier Transform of each sky
response is shown in (d), (e) and (f). A profile of the spatial
response along the $\alpha_x$-axis is also shown. The arrow in the
upper plots indicates the projected wind vector that gives rise to the
temporal filter (thin dark strip) shown in the lower figures. Circles
represent the half-maximum contour of distributions that are
approximately Gaussian.  The shaded region in (e) represents one
channel in the power spectrum, with an effective width
$\Delta\alpha_{\rm chan}$ that is larger than the channel spacing
$\theta_{\rm sweep}^{-1}$ if tapering has been applied to the swept
`slot' in (b). See section 2.3 for a full description of these
figures.\label{response}}
\end{figure*}

A single aperture is chopped between two positions on the sky at equal elevation separated by angle $\theta_{\rm chop}$ and the output is the difference between the two signals. The power gain pattern $G(\theta_x,\theta_y)$ is shown schematically in Fig.~\ref{response}a, with the $\theta_x$ axis defined to be along the chop direction. The individual beams are assumed to be circular Gaussians with FWHM of $\theta_{\rm b}$.

The corresponding spatial filter $\hat{G}^2(\alpha_x,\alpha_y)$, depicted in Fig.~\ref{response}d, is derived using the Fourier Transform relationship in equation~(\ref{G transform}), and is given by
\begin{equation}
\hat{G}^2(\alpha_x,\alpha_y) = \sin^2(\pi\theta_{\rm chop}\alpha_x) 
\exp\left\{-\frac{\pi^2\theta_{\rm b}^2 x^2}{2\log_e 2}\right\}.
\end{equation}
The normalization is such that the maximum value of $\hat{G}(\alpha_x,\alpha_y)$ is unity (Section~\ref{framework}).
The lower half of Fig.~\ref{response}d shows a cross section of the spatial filter along the $\alpha_x$ axis.
The instrument has zero response to brightness corrugations on the sky that have $\alpha_x = n/\theta_{\rm chop}^{-1}, (n = 0,1,2,\ldots)$. It is also insensitive to scales much smaller than the beam size $\theta_{\rm b}$, corresponding to large values of $\alpha$. 

\subsubsection{Swept beam}

The swept beam configuration sweeps the beam from a single aperture back and forth across a strip of sky of width $\theta_{\rm sweep}$ (Fig.~\ref{response}b). The single beam is assumed to be a circular Gaussian with FWHM of $\theta_{\rm b}$, and the $\theta_x$ axis is defined by the sweep direction.
It is assumed that the
data are sampled at the Nyquist rate to generate $N = 2\theta_{\rm
sweep}/\theta_{\rm b}$ samples per sweep. An FFT of these data generate
$N$ real channels (or alternatively $N/2$ complex channels for
$\alpha_x \ge 0$) in the angular wavenumber domain spaced by
$\theta_{\rm sweep}^{-1}$ (dashed lines). The resulting spatial filter is illustrated in Fig.~\ref{response}e. This case is different from
the single chop, since there are $N$ independently determined
quantities, rather than one. Each channel is sensitive to corrugations on the sky with a particular range of wavenumber $\alpha_x$. In practice, data from the swept
`slot' on the sky must to be tapered at the ends to avoid `ringing'
effects in the transform. This gives rise to an effective channel
width $\Delta\alpha$ that is larger than the channel separation, as
indicated by the shaded region in Fig.~\ref{response}e. The beam size
again sets an upper limit on the angular wavenumber to which the
instrument is sensitive.

\subsubsection{Interferometer}

Figure~\ref{response}c and f show the response of an interferometer consisting of two circular apertures producing Gaussian beams with FWHM of $\theta_{\rm b}$. The aperture centers are separated by a baseline of length $B$ and the $\theta_x$ axis is defined to be in the same direction. 
The interferometer responds to a range of spatial wavenumbers centered
on $\alpha_x = \pm B / \lambda$ (Fig~\ref{response}f). The larger the aperture, the larger the range of wavenumbers the instrument is sensitive to. It is assumed that a complex correlator
is used to measure both the sine and cosine components on the sky;
half of the total fluctuation power is present in each component (for
the sake of clarity, Fig.~\ref{response} shows only the cosine
component).  Note that the Gaussian profile would imply that the
interferometer has a finite response to $(\alpha_x, \alpha_y) =
(0,0)$. In fact there must be zero response and the Gaussian
approximation breaks down near the origin, since there is no overlap
between the apertures.

\subsection{Temporal Filtering}

This section derives the form of the temporal filter $\hat{Q}(\alpha_x,\alpha_y)$ of equation~(\ref{filters}). 

The wind vector ${\bf w}$ advects the layer containing the water vapor fluctuations in a horizontal direction, so that the distribution of
fluctuations projected onto the $x-y$ plane appears to move at speed
${\bf w}_{xy} = (w_x, w_y)$, the component of ${\bf w}$ parallel to
the $x-y$ plane. A fluctuation component characterized by wavenumbers$(q_x, q_y)$ sampled along a line of sight parallel to the $z$-axis produces a signal that varies in time with frequency $\nu = w_x q_x + w_y q_y$. Converting to angular wavenumber, 
\begin{equation}
\nu = \frac{\sin\epsilon}{h_{\rm av}} (w_x \alpha_x + w_y \alpha_y).
\end{equation}

Time-averaging the output of an instrument is equivalent to a low pass
filter which rejects signals that are varying rapidly. Since time-averaging is equivalent to convolution of the output time series with a boxcar function, the equivalent frequency response is given by ${\rm sinc}(\pi\nu t_{\rm av})$. Substituting for the frequency $\nu$ we obtain the temporal filter function $\hat{Q}$:
\begin{equation}
\hat{Q}(\alpha_x,\alpha_y) = {\rm sinc}\left(\frac{\pi t_{\rm av}\sin\epsilon}{h_{\rm av}}(w_x \alpha_x + w_y \alpha_y)\right).
\end{equation}
This function is represented schematically by the dark strips in Fig.~\ref{response}d, e and f, perpendicular to the direction of the projected wind vector ${\bf w}_{xy}$.

\subsection{Residual Fluctuation Power}
\label{residual}

The atmospheric fluctuation power remaining at the output of the
instrument is determined by calculating the overlap integral between
the unfiltered atmospheric power (Fig.~\ref{angular_power}) and the
spatial and temporal filtering functions for the instrument as
depicted in Fig.~\ref{response}. This is analogous to the application of a window function for assessing the response of an instrument to the CMB power spectrum, or to the use of the Optical Transfer Function for optical systems. Each of the three instrument configurations is considered below.

\subsubsection{Chopped beam}

The residual level of brightness temperature fluctuations at the output of a chopped beam experiment after time averaging is given by
\begin{eqnarray}
T_{\rm out,rms}^2 &=& 
\int_{-\infty}^{+\infty}\int_{-\infty}^{+\infty} 
\langle\tilde{T}^2_{\rm sky}\rangle\hat{G}^2 \hat{Q}^2 d\alpha_x d\alpha_y \\
&=& \int_{-\infty}^{+\infty}\int_{-\infty}^{+\infty} 
\left\{ 
Ah_{\rm av}^{5/3} (\sin\epsilon)^{-8/3} (\alpha_x^2+\alpha_y^2)^{-11/6}\right\}
\left\{ 
\sin^2(\pi\theta_{\rm chop}\alpha_x)
\right\} \nonumber
\\ && \qquad
\left\{ 
{\rm sinc}^2(\pi t_{\rm av} \frac{\sin\epsilon}{h_{\rm av}}(\alpha_x w_x + \alpha_y w_y))
\right\} d\alpha_x d\alpha_y.
\end{eqnarray}

The integral is dominated by the contribution from close to the origin, in which regime the spatial filter function can be approximated as $\pi^2\theta_{\rm chop}^2\alpha_x^2$. The integral can then be reduced to the following expression:
\begin{equation}
T_{\rm out,rms}^2 \simeq (11.2\cos^2\phi + 16.7\sin^2\phi) 
A h_{\rm av}^2 (\sin\epsilon)^{-3} \theta_{\rm chop}^2 w_{xy}^{-1/3} t_{\rm av}^{-1/3},
\label{chop eq}
\end{equation}
where $\phi$ is the angle between the projected wind vector ${\bf w}_{xy}$ and the $\theta_x$ axis. The expression should apply as long as the 3D turbulence model is a good approximation for $\langle\tilde{T}_{\rm sky}^2\rangle$ (Section~\ref{Kolmogorov}). This requires that the chop angle $\theta_{\rm chop} \ll 2\Delta h \sin\epsilon / h_{\rm av}$ and $w_{xy}t_{\rm av} \ll \Delta h_{\rm av}$. If either constraint is exceeded, equation~(\ref{chop eq}) represents an upper limit, since the regime of `2D turbulence' has reduced power at low angular wavenumbers.  
It is interesting to note that the rms level of the
fluctuations goes down as $t_{\rm av}^{-1/6}$, much more slowly
than the usual $t_{\rm av}^{-1/2}$. This is because the filtered
fluctuations do not have a white noise spectrum; most of the power is concentrated at the low frequencies. 

It is clear that simple chopped observations are very susceptible to
atmospheric fluctuations; in practice more complicated three and four
beam chops have been used to alleviate this.

\subsubsection{Swept beam}

For a channel with $\alpha_x^\prime \gg \Delta\alpha_{\rm chan}$, the atmospheric fluctuation power $\langle\tilde{T}_{\rm sky}^2\rangle$ can be considered approximately constant in the region of overlap between the spatial filter and the temporal filter (Fig.~\ref{response}e). For such a case,
it can be shown that
\begin{eqnarray}
T_{\rm out,rms}^2 &=& 
\int_{-\infty}^{+\infty}\int_{-\infty}^{+\infty} 
\langle\tilde{T}^2_{\rm sky}\rangle\hat{G}^2 \hat{Q}^2 d\alpha_x d\alpha_y \\
&\simeq& 
\left\{Ah_{\rm av}^{5/3}(\sin\epsilon)^{-8/3}
       \left(\frac{\alpha_x^\prime}{\sin\phi}\right)^{-11/3}
\right\}
\left\{\exp\left[-8\ln 2\left(\frac{\alpha_x^\prime}{\alpha_{\rm b} 
       \sin\phi}\right)^2\right]
\right\} \nonumber
\\ && \qquad
\left\{\frac{\Delta\alpha_{\rm chan}}{\sin\phi}
\right\}
\left\{(w_{xy}t_{\rm av})^{-1} \frac{h_{\rm av}}{\sin\epsilon}
\right\}, \label{swept resid}
\end{eqnarray} 
where $\alpha_x^\prime$ is the angular wavenumber of the channel and
$\phi$ is the angle that ${\bf w}_{xy}$ makes with the
$\theta_x$-axis. The first factor is $\langle\tilde{T}_{\rm sky}^2\rangle$ evaluated at the intersection of the channel spatial filter (shaded region in Fig.~\ref{response}e) and the temporal filter (dark strip in Fig.~\ref{response}e). The second factor is the taper imposed by the beam size of the aperture (a wide angle beam is insensitive to high angular wavenumbers). The third factor is the effective length of the temporal filter strip that overlaps with the channel. The fourth factor is the effective width of the temporal filter function. The third and fourth factors represent the area of overlap in Fig.~\ref{response}e; the first and second represent the level of fluctuations in the overlap region. The residual fluctuations are largest when the wind ${\bf w}_{xy}$ is blowing perpendicular to the sweep direction, i.e.~$\phi = 90$~degrees. 

The channels with $\alpha_x^\prime$ close to zero may have a significant
response to the strong peak in $\langle\tilde{T}_{\rm sky}^2\rangle$ near $(\alpha_x,\alpha_y)
= (0,0)$. Careful thought must be given to the tapering (or
`windowing') function applied over the swept slot on the sky, since
this determines the profile of the channel as a function of
$\alpha_x$. Little or no tapering leads to strong sidelobes on the
channel response; too much tapering substantially reduces the resolution of the power spectrum. Specific cases can be computed numerically if the
altitude $h_{\rm av}$ and thickness $\Delta h$ of the layer are known.

\subsubsection{Interferometer}

For an interferometer baseline $B$ that is large compared to the diameter of the antennas, the circles representing the spatial filter response in Fig.~\ref{response}f are widely spaced and the atmospheric power spectral density $\langle\tilde{T}_{\rm sky}^2\rangle$ can be considered constant over these regions. The overlap with the temporal filter (dark strip in Fig.~\ref{response}f) is maximum when the wind blows perpendicular to the baseline. For cases when the wind is close to perpendicular ($\phi > 60^\circ$), the variance of the residual brightness temperature fluctuations at the time-averaged output of the instrument is approximated by
\begin{eqnarray}
T_{\rm out,rms}^2 &=& 
\int_{-\infty}^{+\infty}\int_{-\infty}^{+\infty} 
\langle\tilde{T}^2_{\rm sky}\rangle\hat{G}^2 \hat{Q}^2 d\alpha_x d\alpha_y \label{gsad}\\
&\simeq& 
\left\{Ah_{\rm av}^{5/3}(\sin\epsilon)^{-8/3}
       \left(\frac{B}{\lambda}\sin\phi\right)^{-11/3}
\right\}
\left\{\exp\left[-8\ln 2\left(\frac{B\cos\phi}{\lambda\alpha_{\rm b}}
       \right)^2\right]
\right\} \nonumber
\\ && \qquad
\left\{\frac{\alpha_{\rm b}\sqrt{\pi}}{\sqrt{2\ln2}}
\right\}
\left\{(w_{xy}t_{\rm av})^{-1} \frac{h_{\rm av}}{\sin\epsilon}
\right\}, \label{int resid}
\end{eqnarray} 
The first factor is $\langle\tilde{T}_{\rm sky}^2\rangle$ evaluated at the intersection of the channel spatial filter and the temporal filter (assumes `3D turbulence'). The second factor accounts for the taper of the primary beam. The third factor is the equivalent width of the Gaussian spatial filter along the direction of the temporal filter strip for both positive and negative $\alpha_x$. The fourth factor is the equivalent width of the temporal filter function. 
When the wind blows more parallel to the baseline there is very little overlap between the spatial and temporal filters (Fig.~\ref{response}f) and $T_{\rm out,rms}^2$ is very small compared to the perpendicular case.

If the edge-to-edge separation $\Delta D$ of the apertures is small
compared to their diameter $d$, the circles in Fig.~\ref{response}f are close together, and it is necessary to make a numerical
calculation to account for the change in $\langle\tilde{T}_{\rm sky}^2\rangle$ in the region of overlap between the spatial and temporal filters. For the case where the wind is perpendicular to the baseline, equation~(\ref{gsad}) can be approximated by a one dimensional integral of the atmospheric power law and the spatial filter along the $\alpha_x$ direction, multiplied by the effective width of the temporal filter: 
\begin{equation}
T_{\rm out,rms}^2 \simeq 
\left\{Ah_{\rm av}^{5/3}(\sin\epsilon)^{-8/3}\right\}
\left\{(w_{xy}t_{\rm av})^{-1} \frac{h_{\rm av}}{\sin\epsilon}\right\}
S_{\rm int},
\label{int num resid}
\end{equation}
where
\begin{equation}
S_{\rm int} = \int_{-\infty}^{+\infty} \alpha_x^{-11/3} \hat{G}^2 d\alpha_x.
\end{equation}

\begin{figure*}
\begin{center}
\leavevmode
\epsfxsize=0.75\columnwidth
\epsfbox{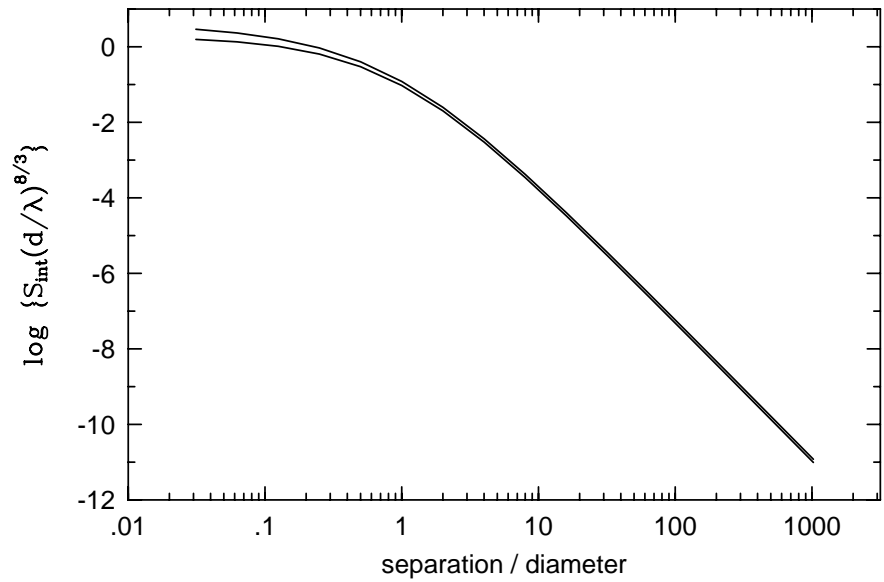}
\end{center}
\caption{Plot of $\log_{10} S_{\rm int} (d/\lambda)^{8/3}$ as a function of 
the ratio of the edge-to-edge separation to the diameter $d$ of the
aperture. The lower of the two lines is for a truncated Gaussian
aperture distribution; the upper is for a corrugated
horn.\label{sep_over_diam}}
\end{figure*}

This integral was evaluated for 
two different types of aperture: (1)~a truncated Gaussian distribution of the electric field strength with the edge cut-off at the -10~dB level; (2)~a truncated Bessel distribution of the electric field, cut off at the first zero. The first is typical for a dish illuminated by a feedhorn, and the second is obtained at the aperture of a corrugated horn. In each case, $\hat{G}$ was calculated by cross-correlating the E-field distribution of two apertures. It was found that the result is almost independent of the transition from
the 3D to the 2D turbulence regime. Also, for a given value of $\Delta D / d$, the integral $S_{\rm int} \propto (d/\lambda)^{-8/3}$ (to a good approximation), i.e.~$S_{\rm int}(d/\lambda)^{8/3}$ is a function of $\Delta D / d$. This is plotted in Fig.~\ref{sep_over_diam}. At large
values of $\Delta D / d$ (where $\Delta D \sim B$), there is a $-11/3$ power-law dependence, as predicted by equation~(\ref{int resid}).

\subsection{The Unfiltered Fluctuation Strength}

The atmospheric brightness fluctuations at a given site and at a given
time can be characterized by four parameters: the wind vector ${\bf
w}$, the altitude of the turbulent layer $h_{\rm av}$, the
fluctuation intensity $A$, and the thickness of the layer $\Delta h$.
Of particular importance is the combination $A h_{\rm av}^{8/3}$,
which is the measure of fluctuation `strength' relevant to the swept 
beam and interferometric observations of the CMB (eqs.~[\ref{swept resid}] and [\ref{int resid}]).

In the next sections, data from the South Pole and Chile are analyzed to 
constrain these parameters. The models developed above are then used to 
estimate the sky brightness fluctuations that would be expected for an 
interferometer located at these sites.

\section{The South Pole}
\label{sect:sp}
\label{python}

The South Pole has been chosen as a site for several CMB
anisotropy experiments over the past decade (see for example
Meinhold \& Lubin 1991, Tucker et al. 1993, Dragovan et al. 1994,
Platt et al. 1997). It is high (2800 m), extremely cold and dry, and
is situated on an expansive ice sheet, with the surface wind
dominated by weak katabatic airflow from higher terrain several
hundred kilometers away to grid northeast (see discussion in King \&
Turner 1997). We characterize atmospheric brightness fluctuations at
the South Pole using data from the Python telescope, a swept beam
CMB experiment, obtained during the austral summer 1996-1997.

\subsection{The Python Experiment}
\label{sect:sp:exp}

The Python telescope, in its configuration for the 1996-1997
season, employs a dual feed 40~GHz HEMT based receiver. The receiver
has two corrugated feeds separated by $2.75^\circ$ on the sky, with 
separate RF chains, HEMT amplifiers, and backend signal
processing. The post-detector output is AC coupled with a cut-off
frequency of 1~Hz, and incorporates a low-pass anti-alias filter which
attenuates the signal at frequencies above 100~Hz.  The RF signal is
separated into two frequency bands, centered at 39~GHz and 41.5~GHz with
bandwidths of approximately 2~GHz and 5~GHz, respectively. Data from
the two bands are combined for atmospheric analysis.

The feeds are in the focal plane of a 0.8~m off-axis paraboloidal
mirror, resulting in two $1.1^\circ$ beams on the sky which are
swept through $10^\circ$ at constant elevation at a rate of 5.1~Hz by
a large vertical flat mirror. The two beams are at the same elevation
and their sweeps partially overlap on the sky.  Beam spillover is
reflected to the sky by two sets of shields, one set fixed to the
tracking telescope, and a set of larger stationary ground shields
which also shield the telescope structure from the sun and any local
sources of interference. Data are taken for $\sim30$~s while sweeping
and tracking a central position on the sky; the telescope is then
slewed to another position a few degrees away. Between 5 and 13
pointings are stored in one data file, representing 5 to 10 minutes of
observing time. Data were taken over 80\% of the period from early
December 1996 through early February 1997.  

\subsection{Data Analysis Technique}
\label{sect:sp:analysis}

In order to differentiate atmospheric fluctuation power from
instrument noise, the covariance of the data from the two beams is taken
for the portion of each sweep in which their positions overlap on the
sky, approximately $6^\circ$. Atmospheric brightness fluctuations are
correlated between the two beams, while most of the instrument noise is
uncorrelated.  Thus the signal-to-noise of the correlated fluctuation
power can be increased by averaging the covariance over many
sweeps on the sky, allowing the atmospheric brightness 
fluctuation power to be estimated during stable periods when the system
is receiver noise limited. The mean covariance represents the mean
`snapshot' fluctuation power in the $6^\circ$ sweep, regardless of the 
number of sweeps that are subsequently averaged together. 

Several instrumental effects are accounted for when estimating
the atmospheric brightness fluctuation power:

\begin{enumerate}

\item{The fluctuation power is corrected for the effect of the
anti-alias filter roll-off and backend electronics delay.}

\item{Correlated 60 Hz line noise is removed from the data.}

\item{An offset dependent on the position of the sweeping mirror
is correlated between the two beams. This is removed by subtracting
the component of the signal that is constant on the sky over
multiple pointings.}
\end{enumerate}

We determine the effect of the 60~Hz line noise and
stationary signal removal techniques on the true atmospheric signal
by examining their effect on the data during periods when the data
are dominated by atmospheric fluctuations. We calculate that the
atmospheric power should be increased by 30\% to compensate for the
combined effect of these removal techniques and instrumental effects. There
is an additional correlated signal due to the stationary ground
shield and due to the CMB itself, which is not removed with these
techniques. Therefore, the quartiles which we report should be taken
as an upper limit of the true atmospheric signal.

\subsection{South Pole Atmospheric Fluctuation Data}
\label{sect:sp:data}

\begin{figure*}
\begin{center}
\leavevmode
\epsfxsize=0.75\columnwidth
\epsfbox{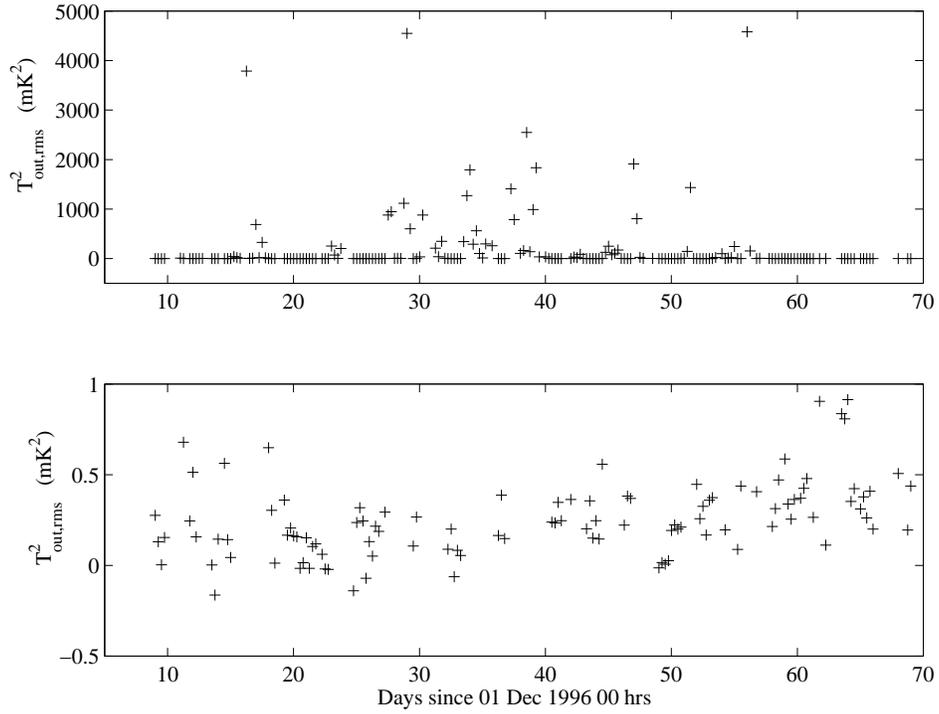}
\end{center}
\caption{The brightness fluctuation power, in mK$^2$, for 2 months of Python
observations at the South Pole. The measured fluctuation power is for
a $6^\circ$ sweep on the sky, at a mean elevation angle of
$49^\circ$. The data have been binned in 6 hour intervals to reduce
instrument noise. In the lower plot, the scale of the vertical axis is
enlarged about zero to show the fluctuation power during stable
periods. Even after binning into 6 hour intervals, the variability in
the fluctuation power during stable periods is consistent with
instrument noise.\label{fig:covariance}}
\end{figure*}

>From the Python data, we determine the South Pole brightness
fluctuation power for a 6$^\circ$ sweep at a mean elevation angle of
$49^\circ$ over 2 months of the austral summer
(Fig.~\ref{fig:covariance}). The atmospheric fluctuations at the South
Pole are bimodal in nature, with long periods of high stability broken
by periods of high atmospheric fluctuation power. An examination of
the meteorolgical records shows that the latter are always associated
with at least partial cloud cover.  The infrequent periods of high
atmospheric fluctuation power are correlated with a (grid) westerly
shift in the wind direction, and are weakly correlated with increased
windspeed and precipitable water vapor content. These correlations
indicate that the periods of high atmospheric fluctuation power are
associated, at least in part, with synoptically forced moist air from
West Antarctica, a condition which occurs infrequently at the pole
(Hogan et al. 1982). Variability in the fluctuation power during
periods of low atmospheric fluctuation power is consistent with
instrument noise.

\begin{figure*}
\begin{center}
\leavevmode
\epsfxsize=0.75\columnwidth
\epsfbox{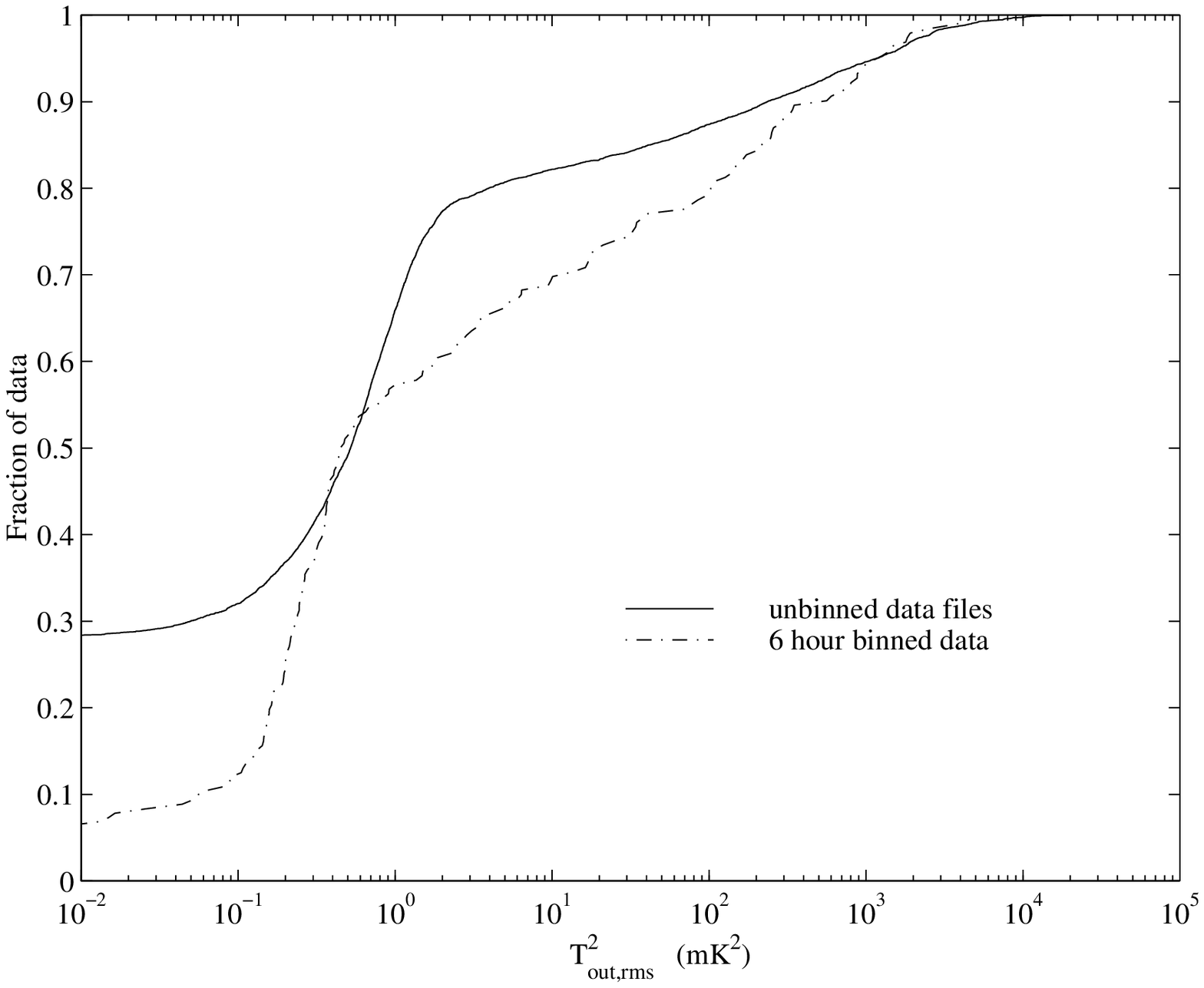}
\end{center}
\caption{The cumulative distribution function of brightness
fluctuation power for the Python data shown in
Fig.~\ref{fig:covariance}.  Two distributions are shown, one for
unbinned data files taken over a few minutes (solid line), and one for
the data binned in 6 hour intervals (dash-dot line). The unbinned data
files are adequate to determine the 50\% and 75\% quartile values, but the 6
hour binned cumulative distribution function is needed to achieve
adequate signal-to-noise to estimate the 25\% quartile value. The increased
fraction of high fluctuation power data in the 6 hour binned
cumulative distribution function results from binning brief periods of
high fluctuation power together with periods of low fluctuation
power. The quartile values are \{0.20, 0.51, 1.62\}~mK$^2$ (before
compensating for instrumental effects).\label{fig:cdf}}
\end{figure*}

From these data we construct a cumulative distribution function for
the brightness fluctuation power (Fig.~\ref{fig:cdf}), and derive
quartile values for brightness fluctuation power over the 2 month time
period during which the data were taken. The cumulative distribution
function for the data files, in which the covariance has been averaged
for only a few minutes, is adequate for deriving the 50\% and 75\%
quartile values. However, instrument noise dominates the distribution
function at the 25\% quartile level; therefore data that are binned in
6 hour intervals are used to derive the 25\% quartile value. The
quartile values for the 2 months of the austral summer 1996-1997 are
\{0.20, 0.51, 1.62\}~mK$^2$, where the brackets are used to denote the
25\%, 50\% and 75\% quartile values. To compensate for instrumental
filtering effects described above, we increase these values by 30\% to
\{0.27, 0.68, 2.15\}~mK$^2$. The Python experiment was not operated
during the austral winter, so data are not available for this
period. However, precipitable water vapor and sky opacity quartile
values are lower in the winter months (Chamberlin et al. 1997). It is
therefore likely that the atmospheric stability improves during the
austral winter as well.

The sharp roll-off of the Python primary beam spatial filter (see
Fig.~\ref{response}e) prevents an accurate determination of the
underlying atmospheric angular power spectrum.  The angular power
spectrum for data that have not been tapered at the edges of the sweep
exhibits power proportional to the inverse square of the angular
wavenumber. This power law dependence is due to power from low angular
wavenumbers (large angular scales) leaking into higher angular
wavenumber channels. The observations demonstrate that it is desirable
to taper CMB data to reduce contamination from the high atmospheric
fluctuation power on large angular scales.

\subsection{Estimating the fluctuation intensity}

The value of $A h_{\rm av}^{5/3}$ (eq.~[\ref{sdge}]) appropriate for
the South Pole can be estimated from the rms of the Python
measurements.  Model atmospheric fluctuation power spectra for layers
of turbulence seen at an elevation of $49^\circ$ were computed using a
full three dimensional integration (similar to Lay~1997). These were
then multiplied by the Python spatial filter function $\hat{G}^2$ (the
sum of all but the DC channel) and integrated to give a value for
$\Delta T_{\rm out,rms}^2$. The model value for $A h_{\rm av}^{5/3}$
was then scaled so that the model $\Delta T_{\rm out,rms}^2$ matched
the measured value. Note that since we are estimating the level of rms
fluctuation level from `snaphots' of the atmosphere, there is no
correction needed for time averaging, i.e.~the temporal filter
$\hat{Q}^2=1$ in equation~\ref{filters}.

The result is a function of the ratio of the layer thickness to the
average altitude of the layer, $\Delta h / h_{\rm av}$.  For $\Delta h
/ h_{\rm av} = 1$, it was found that $A h_{\rm av}^{5/3} = 0.99 \Delta
T_{\rm out,rms}^2$; for $\Delta h / h_{\rm av} = 2$, 0.5 and 0.25, the
scale factor changed to 0.89, 1.13 and 1.36, respectively. Therefore
the ratio does not have a major effect on the inferred value of $A
h_{\rm av}^{5/3}$.  For $\Delta h = h_{\rm av} = 500$~m, the measured
quartile values give $A h_{\rm av}^{5/3} = 0.99\Delta T_{\rm atm}^2 =
\{0.27, 0.67, 2.13\}$~mK$^2$. 
Corresponding values for the quantity $A h_{\rm av}^{8/3} / ({\rm mK^2
m})$ (which determines the residual level of fluctuations for swept
beam and interferometer experiments -- see eqs.~[\ref{swept resid}] and
[\ref{int resid}]) are tabulated in Table~\ref{conditions} for three
different values of $h_{\rm av}$.  These numbers are appropriate for
an observing frequency of 40~GHz.  The brightness temperature of
fluctuations due to water vapor scales approximately as the square of
the observing frequency (except close to the line centers at 22~GHz
and 183~GHz), so that $A h_{\rm av}^{8/3}$ should be scaled as observing
frequency to the fourth power.


\begin{table}[thb]
\begin{center}
\begin{tabular}{cccccccc}
\hline\hline
  & \multicolumn{3}{c}{South Pole $h_{\rm av}$} &
  & \multicolumn{3}{c}{Chile $h_{\rm av}$} \\

Quartile\quad & 500~m & 1000~m & 2000~m & \quad & 500~m & 1000~m & 2000~m \\
\hline
25\% &  150 &  340 &  800 & &    1500 &    9400 &    60,000 \\
50\% &  370 &  840 & 2000 & &    7000 &  44,000 &   290,000 \\
75\% & 1200 & 2700 & 6400 & &  29,000 & 190,000 & 1,200,000 \\
\hline
\end{tabular} 
\end{center}
\caption{Values of $A h_{\rm av}^{8/3}$ in units of mK$^2$~m (see eqs.~[\ref{swept resid}] and [\ref{int resid}]) for the
South Pole as estimated from Python V data (section~\ref{python}), and for
the Atacama Desert in Chile estimated from phase monitor data
(section~\ref{chile}). The altitude $h_{\rm av}$ of the turbulent layer is not known. The Chile numbers have a 50\% uncertainty associated
with the conversion from refractive index to brightness
temperature. Values are appropriate for 40~GHz, but can be scaled to
other frequencies based on the emissivity spectrum of water vapor.}%
\label{conditions}
\end{table}

\subsection{Altitude of the fluctuations}

\begin{figure*}
\begin{center}
\leavevmode
\epsfxsize=0.75\columnwidth
\epsfbox{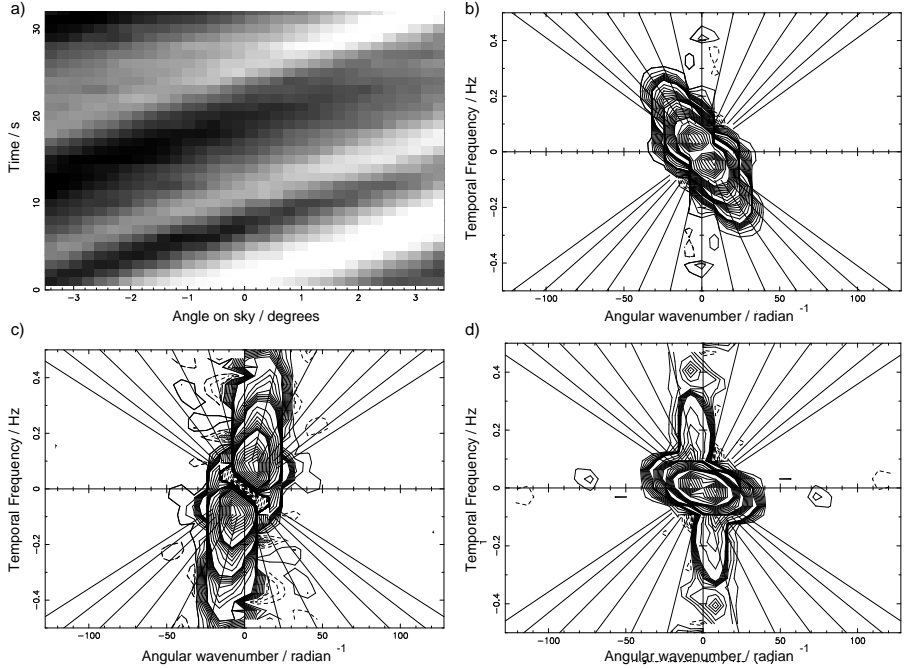}
\end{center}
\caption{a)~Time--angle plot for 30~s of Python~V data during bad 
weather, illustrating the striping that results when the wind blows
blobs of water vapor along the sweep direction. b)~Power spectrum for
a 1~hour period which includes the sample shown in (a). Contours are
relative to maximum (0.9, 0.8, 0.7, ..., 0.2, 0.1, 0.09, 0.08,
etc.). Note that the power is concentrated in a line that is
perpendicular to the striping in (a). The radial lines correspond to
different altitudes, based on the windspeed measured by radiosonde
launches. The y-axis corresponds to ground level, and subsequently
shallower lines represent 500, 1000, 1500, 2000, 2500 and 3000~m,
respectively.  The fluctuations in (a) are at roughly 1500~m above
ground. c)~Another example showing fluctuations at less than
500~m. d)~Two fluctuation components; one very low, the other very
high.\label{time-angle}}
\end{figure*}

By combining the Python data with radiosonde wind measurements, it is
possible to determine the altitude of the fluctuations during periods
of bad weather. This is illustrated by Fig.~\ref{time-angle}, which
shows the measured emission as a function of angular position
$\theta_x$ on the sky and time $t$. The plot represents an interval of
30~s, at a time when the wind was blowing parallel to the sweep
direction. The stripes are produced by blobs of water vapor moving
from left to right; the diagram shows that a blob moves through an
angle of $7^\circ$ in about 13~s.  A $7^\circ$ angular distance
corresponds to a physical length of $0.12 h_{\rm av} / \sin
(49^\circ)$, where $h_{\rm av}$ is the average altitude of the
fluctuation. The radiosonde launched 2 hours after this dataset
indicated a fairly uniform wind speed of $w = $16~m~s$^{-1}$ for the
lower 2~km, so a blob is expected to move 208~m in 13~s. Solving for
$h_{\rm av}$ gives 1300~m.  The slope of the stripes is proportional
to $h_{\rm av}/w$.

In order to average the data together from many 30-s intervals, we
compute the power spectrum for each time--angle plot and average those
together. The power spectrum for a 30-s interval is calculated by
computing the Fast Fourier Transform of the time--angle data from each
of the two feedhorns (using a Hann taper to minimize sidelobes), and
then calculating the covariance between the two transformed
datasets. Figure~\ref{time-angle}b shows the power spectrum averaged
over one hour of data (containing the 30-s interval shown in
Fig.~\ref{time-angle}a). The power is distributed along a radial line,
perpendicular to the striping. The gradient of this line is
proportional to $w/h_{\rm av}$; the overlaid radial lines represent
(from vertical) altitudes of 0, 500, 1000, 1500, 2000, 2500 and
3000~m, calibrated using $w = $16~m~s$^{-1}$. The fluctuations have
$h_{\rm av} \sim 1300$~m. The physical periodicity is given by
$w/\nu$; e.g.~$\nu = 0.1$~Hz corresponds to a fluctuation with period
160~m. There is little power present at the origin because the DC
level was removed from the time--angle plane before the Hann taper
and Fourier transform were applied. The contours appear to fall off 
faster than would be expected for Kolmogorov turbulence. This is partly
due to the effect of the primary beam taper in the angular wavenumber
direction, which can be represented by a Gaussian centered on zero with a
FWHM size of about 33 rad$^{-1}$, but may also indicate that these
bad weather fluctuations do not follow a Kolmogorov power law.
 
Average power spectra were computed for all the hours when the wind
was parallel to the sweep direction (approximately every 12 hours). Two
more examples during periods of bad weather are shown in
Fig.~\ref{time-angle}c and d. The first indicates fluctuations at an
altitude of $\sim 500$~m; the second shows two components: one at
$\sim 300$~m and another at much higher altitude
($>3$~km). Unfortunately it was not possible to detect structure in
the power spectrum during stable periods; the emission is too weak.

In most cases that were measured, the altitude determined for the
strong fluctuations, which varies from 300~m to well over 3~km, agreed
well with the altitude at which the relative humidity was a maximum
(measured by radiosonde launches). This strengthens the case for the
connection between clouds and strong fluctuations, and may explain the
possible non-Kolmogorov nature of the power spectrum. In the cases of
high altitude turbulence ($h_{\rm av}>3$~km), however, there was no
corresponding maximum in the relative humidity and there is generally
little water vapor present in the atmosphere. Another mechanism must
be at work in these cases.

\section{The Atacama Desert in Chile}
\label{chile}

The Atacama Desert in Northern Chile is extremely dry, and is the
proposed location for the next generation of large millimeter-wave
arrays, as well as the Cosmic Background Interferometer experiment.
The Mobile Anisotropy Telescope has been deployed there for two
seasons during 1997-1998 (Torbet et al.\ 1999, Miller et al.\
1999). Monitoring of the atmospheric stability at Cerro Chajnantor
(the proposed site for the Atacama Large Millimeter Array) has been
underway for over 4 years, using a site test interferometer (Radford
et al.~1996).

\subsection{The site test interferometer}

The interferometer used in Chile consists of two dishes 1.8~m in
diameter, separated by an East-West baseline of 300~m, that observe
the 11~GHz CW beacon from a geostationary communications satellite at
an elevation of 36$^\circ$ and an azimuth of 65$^\circ$. The
instrument measures the phase difference between the two received
signals, which depends on the difference in the electrical path
lengths along the two lines of sight to the satellite. The random
component of the fluctuations is dominated by the non-uniform
distribution of water vapor in the troposphere being blown over the
interferometer, since water vapor has a high refractive index compared
to dry air. Quartile values for the rms difference in path length over 1996
are \{136, 300, 624\}~$\mu$m. The
data processing is described by Holdaway et al.\ (1995). These values 
have not been scaled to zenith.

Water vapor is the principal source of both path length fluctuations
and brightness temperature fluctuations. We wish to estimate the
latter from a measurement of the former, which requires a conversion
factor. This is described in \S\ref{conversion}. It is also necessary
to determine $A$ (the intensity of fluctuations integrated through the
atmosphere) from the interferometer measurements. This requires a model,
and is described in \S\ref{column density}.

\subsection{Converting rms path to brightness temperature}
\label{conversion}

The millimeter and submillimeter absorption spectrum of water vapor is
dominated by a series of rotational line transitions, the lowest of
which are at 22~GHz and 183~GHz. At frequencies close to the line
centers, the optical depth $\tau$ for a given amount of precipitable
water vapor (PWV: the depth of liquid water obtained if all the vapor
is condensed) is well-determined, but between the lines experiments
have shown that the absorption is higher than expected from theory
(e.g.~Waters 1976, Sutton and Hueckstaedt 1996). The reason for this
is still not clear, although several hypotheses have been advanced. At
40~GHz the theoretical absorption from lines is 0.012 per centimeter
PWV. The Gaut and Reifenstein (1971) model for the `continuum'
absorption increases this to 0.03 cm$^{-1}$, and is believed to be a
reasonable estimate at the frequencies of interest here (Sutton and
Hueckstaedt 1996). If the physical temperature of the atmosphere is
250~K, then 1~cm PWV has a brightness temperature of $0.03 \times
250$~K$ = 7.5$~K. The excess path resulting from 1~cm of water vapor
at 250~K is approximately 7~cm (Thompson, Moran \& Swenson 1994), so
that 1~cm of excess path corresponds to a brightness temperature of
1.1~K, and 1~mK brightness corresponds to 9~$\mu$m excess path. This
conversion should be considered an estimate with an uncertainty of
order 50\%. A hard upper limit of 22~$\mu$m~mK$^{-1}$ is set by the
contribution from line emission only (no continuum term).

Using the value of 9~$\mu$m~mK$^{-1}$, the rms path fluctuation 
quartile values are mapped to brightness temperature fluctuations of \{16, 34, 69\}~mK at 40~GHz.

\subsection{Estimating $A$}
\label{column density}

Since the phase monitor measures the path difference between two lines
of sight, its spatial filtering properties are analogous to the
chopping instrument depicted in Fig.~\ref{response}a and d. The main
difference is that the lines of sight are parallel to one another
through the atmosphere, rather than diverging from the point of
observation. As discussed in \S\ref{residual}, calculation of the
residual fluctuation power after spatial and temporal filtering for
this configuration requires a model that includes the thickness of the
layer containing the fluctuations and the detailed geometry of the
observations. The relevant analysis can be found in Lay~(1997).

The input model parameters are: baseline 300~m East-West; elevation
$36^\circ$; azimuth $65^\circ$; layer thickness 500~m. The windspeed
and altitude of the layer are not needed for calculation of $A$. To
obtain rms brightness temperatures of $\{16, 34, 69\}$~mK with the
above model parameters requires that $A =
\{9.4\times 10^{-5}, 4.4\times 10^{-4}, 
1.9\times 10^{-3}\}$~mK$^2$~m$^{-5/3}$.  The layer thickness has only
a small effect on the calculation; if instead the layer is actually
2~km thick, then $A = \{8.0\times 10^{-5}, 3.7\times 10^{-4},
1.6\times 10^{-3}\}$~mK$^2$~m$^{-5/3}$.

The average altitude of the turbulent layer is much more important,
but is not known. The quantity $A h_{\rm av}^{8/3} / ({\rm mK^2 m})$
is tabulated in Table~\ref{conditions} for different values of $h_{\rm
av}$ and $\Delta h = 500$~m. This is the value that determines the
residual level of fluctuations in a CMB experiment
(eq.~[\ref{int resid}]).

\section{Example}
\label{example}

The analysis and results presented above are applied to calculate the
contribution from atmospheric brightness fluctuations to the noise
measured by a small interferometer, of the kind that might be used for
measurements of the CMB anisotropy.

We specify a field of view with FWHM of $\theta_{\rm b} = 3^\circ =
0.053$~rad at a frequency of 40~GHz ($\lambda = 0.75$~cm). The
diameter of the aperture required for a Gaussian distribution
truncated at the $-$10~dB level is $d = 1.15 \lambda / \theta_{\rm b} =
16.3$~cm. For a corrugated horn, the diameter is $d = 1.32 \lambda /
\theta_{\rm b} = 18.7$~cm, and it is this case that we now investigate 
further.

The total thermal noise (for the combined sine and cosine components, no atmospheric fluctuations) for a
2-element interferometer with system temperature $T_{\rm sys}$ and bandwidth
$\Delta\nu$ is given by (see Thompson, Moran \& Swenson 1994)
\begin{equation}
T_{\rm therm,rms} = 0.28 \left(\frac{T_{\rm sys}}{20~{\rm K}}\right) 
                         \left(\frac{\Delta\nu}{5~{\rm GHz}}\right)^{-1/2} 
                         \left(\frac{t_{\rm av}}{\rm s}\right)^{-1/2}
                         \:{\rm mK}.
\end{equation}
The rms atmospheric contribution is given by the square root of equation~(\ref{int num resid}):
\begin{equation}
T_{\rm out,rms} 
	= \left(A h_{\rm av}^{8/3} \right)^{1/2} (\sin\epsilon)^{-11/6} \,         w_{xy}^{-1/2} \, t_{\rm av}^{-1/2} S_{\rm int}^{1/2} .
\label{hjrs}
\end{equation}
The value of $S_{\rm int}$ is determined from
Figure~\ref{sep_over_diam}, and depends on the separation of the apertures. The value of $A
h_{\rm av}^{8/3}$ depends on the conditions at the site, as shown in
Table~\ref{conditions}, as does the projected windspeed $w_{xy}$.

An ideal site should not significantly compromise the sensitivity of
the experiment. We can determine the range of $Ah_{\rm av}^{8/3}$ for
which $T_{\rm out,rms} < T_{\rm therm,rms}$. From
Fig.~\ref{sep_over_diam}, the maximum value of $\log S_{\rm
int}(d/\lambda)^{8/3}$ is approximately $+0.3$, which applies when the
wind is perpendicular to the baseline for an interferometer with
corrugated horns that are very close together. In this example, $d/\lambda = 24.9$, which implies $S_{\rm int} \leq 3.8\times 10^{-4}$.
Substituting these values into equation~(\ref{hjrs}), along with a representative elevation of $60^\circ$ and a projected windspeed of 10~m~s$^{-1}$, we obtain
\begin{equation}
T_{\rm out,rms} \leq 8.0 \times 10^{-3} 
	\left(A h_{\rm av}^{8/3} \right)^{1/2} 
	t_{\rm av}^{-1/2} {\rm m}^{1/2} {\rm s}^{-1/2}. 
\end{equation}
For this to be less than the thermal noise contribution we require
$Ah_{\rm av}^{8/3} < 1.2\times 10^3$~mk$^2$~m$^{-1}$.  Reference to
Table~\ref{conditions} indicates that the South Pole should satisfy
this requirement most of the time during the summer months, unless 
the fluctuations are higher
than 2~km above the ground. 
At the Chile site, it appears that the instrument noise for this configuration would be dominated by atmospheric fluctuation power, regardless of the altitude of the fluctuations. It is important, however, to realize that this example was calculated
for the extreme case of the wind blowing perpendicular to a baseline
for which the apertures are almost touching. The situation is 
improved by separating the apertures to observe smaller angular scales
(Fig.~\ref{sep_over_diam}), or when the wind blows parallel to the
baseline (Fig.~\ref{response}f).

\section{Summary}

\begin{enumerate}

\item
The impact of non-uniform emission from the atmosphere on measurements 
of the CMB is assessed. Chopped beam, swept
beam and interferometric measurement schemes are analyzed.

\item
The analysis is based on a model where the atmospheric fluctuations
are confined to a turbulent layer. Data from phase monitors indicate
that there is a transition from a three- to a two-dimensional regime
on scales comparable with the thickness of the layer. The outer
correlation scale of fluctuations is much larger than the layer
thickness. Previous analyses have assumed a much smaller outer scale
length and therefore underestimate the level of fluctuations on large
scales. The impact of atmospheric fluctuations is assessed by
considering the instruments as a combination of spatial and temporal
filters that act on the power spectrum of the turbulence.

\item
Data from the Python V experiment during summer at the South Pole are
analyzed to determine the level of fluctuations. The distribution is
bimodal, with stable conditions for 75\% of the time and much stronger
fluctuations 25\% of the time.

\item 
During normal stable conditions the rms brightness temperature
variation across a $6^\circ$ strip was less than 1~mK, at a frequency of
40~GHz. It was not possible to determine either the average altitude
or the power spectrum of these very weak fluctuations.

\item 
The bad weather fluctuations have rms brightness temperature
variations of up to 100~mK across a $6^\circ$ strip of sky, at a
frequency of 40~GHz. They are only present when there is at least
partial cloud cover. The windspeed as a function of altitude was used
to determine the average altitude of the fluctuations. This varied
from 500~m to over 4~km, and in most cases corresponded to the
altitude at which the relative humidity was a maximum. The power
spectrum measured for these bad weather fluctuations appears to fall
off faster than the predicted Kolmogorov power law, although this may
be due to the taper imposed by the primary beam. We conclude that
these strong fluctuations are probably associated in some way with
cloud activity.

\item
Path fluctuation data from a satellite phase monitor located in the
Atacama Desert in Chile were used to estimate the corresponding level
of brightness temperature fluctuations at 40~GHz for that site.
Although this conversion is uncertain by a factor of 2, and the result
depends strongly on the unknown altitude of the turbulent layer, the
two months of South Pole data indicate a significantly lower level of 
fluctuations compared to the Chile site, as shown in 
Table~\ref{conditions}.

\item
The theoretical analysis and experimental measurements were combined
to predict the residual atmospheric noise that would be present at the
output of a small interferometer with antennas of diameter $\sim
25\lambda$ that could be used to measure CMB anisotropy.  It was found
that the atmospheric contribution is likely to be small compared to
the thermal noise when the apertures are well-separated, but that the
atmosphere can dominate in cases where the edges of the apertures are
separated by only a few wavelengths. For the latter case, a good site,
such as the South Pole is critical.

\item
The analysis shows that the average altitude of the turbulent layer 
has a big impact on the suitability of a site for CMB
measurements. This parameter is not well constrained by the data from
either site. Measurement of this altitude should be a priority for future
site testing experiments.

\end{enumerate}

\section*{Acknowledgments}

The authors wish to express their gratitude to all of the members of
the Python V observing team: S. R. Platt, M. Dragovan, G. Novak,
J. B. Peterson, D. L. Alvarez, J. E. Carlstrom, J. L. Dotson,
G. Griffin, W. L. Holzapfel, J. Kovac, K. Miller, M. Newcomb, and
T. Renbarger, who provided us with the Python V data.  We thank John
Kovac, John Carlstrom and Bill Holzapfel for their insights and
helpful discussions, and Greg Griffin and Steve Platt for providing
assistance with data and code. We thank the South Pole meteorology
office for providing weather data.  This research was supported in
part by the National Science Foundation under a cooperative agreement
with the Center for Astrophysical Research in Antarctica (CARA), grant
NSF OPP 89-20223.  CARA is a National Science Foundation Science and
Technology Center.

\parindent=0in

\section*{References}

Andreani, P., Dall'oglio, G., Martinis, L., Piccirillo, L., \& Rossi, L. 
1990, Infrared Phys., 30, 479 

Armstrong, J. W., \& Sramek, R. A., 1982, Radio Science 17, 1579

Chamberlin, R. A., Lane, A. P., \& Stark, A. A. 1997, ApJ, 476, 428

Church, S. E. 1995, MNRAS, 272, 551

Coble, K., Dragovan, M., Kovac, J., Halverson, N. W., Holzapfel, W. L., 
Knox, L., Dodelson, S., Ganga, K., Alvarez, D., Peterson, J. B., Griffin,
G., Newcomb, M., Miller, K., Platt, S. R., Novak, G. 1999, ApJ, 519, L5 

Coulman, C. E., \& Vernin, J., 1991, Applied Optics 30, 118 

Davies, R. D., Gutierrez, C. M., Hopkins, J., Melhuish, S. J., Watson, R.
A., Hoyland, R. J., Rebolo, R., Lasenby, A. N., \& Hancock, S. 1996, MNRAS,
278, 883

Dragovan, M., Ruhl, J., Novak, G., Platt, S., Crone, B.,
Pernic, R., \& Peterson, J. 1994, ApJ, 427, L67

Gaut, N. E., \& Reifenstein, E. C. III, 1971, Environmental Res. and
Tech. Rep. No. 13 (Lexington, Mass.)

Halverson, N. W., Carlstrom, J. E., Dragovan, M., Holzapfel, W. L., 
Kovac, J. 1998, In: Phillips, T. G. (Ed.), Advanced Technology
MMW, Radio, and Terahertz Telescopes, Proc. SPIE, 3357, p.~416

Hogan, A., Barnard, S., Samson, J., \& Winters, W. 1982, J. Geophys. Res.,
87, 4287

Holdaway, M. A., Radford, S. J. E., Owen, F. N., \& Foster, S. M. 1995,
Millimeter Array Technical Memo No. 129 

Jones, M. E., \& Scott, P. F. 1998, The Very Small Array: Status Report, In: Tran Thanh Van, J., Giraud-Heraud, Y., Bouchet, F., Damour, T. \& Mellier, Y. (eds.), Fundamental Parameters in Cosmology, Proceedings of the 33rd
Recontres de Moriond, p.~233

King, J. C., \& Turner, J. 1997, Antarctic Meteorology and Climatology
(Cambridge: Cambridge University Press), 92

Kovac, J., Dragovan, M., Schleuning, D. A., Alvarez, D., Peterson, J. B.,
Miller, K., Platt, S. R., \& Novak, G. 2000 in preparation

Lay, O. P. 1997, A\&AS, 122, 535

Leitch, E. M., Readhead, A. C. S., Pearson, T. J., Myers, S. T., \& Gulkis,
S. 1998, ApJ, submitted (astro-ph/9807312)

Masson, C. R., 1994, Atmospheric Effects and Calibrations, In: 
Ishiguro, M. \&  Welch, W. J. (eds.),
Astronomy with Millimeter and Submillimeter Wave Interferometry, ASP 
Conference Series Vol. 59, p.~87

Meinhold, P., \& Lubin, P. 1991, ApJ, 370, L11

Miller, A. D., Caldwell, R., Devlin, M. J., Dorwart, W. B., Herbig, T.,
Nolta, M. R., Page, L. A., Puchalla, J., Torbet, E., Tran, H. T. 1999,
ApJ, 524, L1

Netterfield, C. B., Devlin, M. J., Jarolik, N., Page, L.,
Wollack, E. J. 1997, ApJ, 474, 47

Platt, S. R., Kovac, J., Dragovan, M., Peterson, J. B.,
\& Ruhl, J. E. 1997, ApJ, 475, L1

Radford, S. J. E., Reiland, G., \& Shillue, B. 1996, PASP, 108, 441

Ruhl, J. E., Dragovan, M., Platt, S. R., Kovac, J., \& Novak, G. 1995, ApJ,
453, L1

Scott, P. F., Saunders, R., Pooley, G., O'Sullivan, C.,
Lasenby, A. N., Jones, M., Hobson, M. P., Duffett-Smith, P. J.,
\& Baker, J. 1996, ApJ, 461, L1

Sutton, E. C., \& Hueckstaedt, R. M. 1996, A\&AS, 119, 559

Tatarskii, V. I., 1961, Wave Propagation in a Turbulent Medium, Dover: New York

Tegmark, M., \& Efstathiou, G. 1996 MNRAS, 281, 1297

Thompson, A. R., Moran, J. M., \& Swenson, G., 1994, 
Interferometry and Synthesis in Radio Astronomy, Krieger

Torbet, E., Devlin, M. J., Dorwart, W. B., Herbig, T., Miller, A. D.,
Nolta, M. R., Page, L., Puchalla, J., Tran, H. T. 1999, ApJ, 521, L79

Treuhaft, R. N., \& Lanyi, G. E. 1987, Radio Sci., 22 , 251

Tucker, G. S., Griffin, G. S., Nguyen, H. T., \& Peterson, J. B.
1993, ApJ, 419, L45

Waters, J. W. 1976, Methods of Experimental Physics, Vol.~12B 
(M. L. Meeks, ed.), pp. 142-176

Wright, M. C. H. 1996, PASP, 108, 520

\end{document}